# Junctionless dual-gate electrostatic modulation of self-aligned oxide channels by chitosan-based proton conductors

Wei Dou[1], Qing Wan[1, a)], Jie Jiang[1, 2)], Liqiang Zhu[1)], and Qing Zhang[2, b)]

1) Ningbo Institute of Materials Technology and Engineering, Chinese Academy of Sciences, Ningbo 315201, People's Republic of China.
2) Microelectronics Centre, School of Electrical and Electronic Engineering, Nanyang Technological University, 639798, Singapore.

*Abstract*

Dual-gate electrostatic modulation gives an attractive approach for transistors performance improvement, threshold voltage ($V_{th}$) and operation mode modulation, which is favorable for chemical sensor and logic applications. Here, a self-aligned junctionless semiconducting oxide channels are dual-gate electrostatic modulated by solution-processed chitosan-based proton conductors on paper substrates. The low-voltage junctionless paper transistors can be effectively tuned from depletion mode to enhancement mode by the second in-plane gate. OR logic gate was experimentally demonstrated on such dual-in-plane gate junctionless transistors. Such dual-gate organic/inorganic hybrid paper transistors are promising for portable paper electronics.

a) Corresponding authors. Qing Wan, E-mail address: wanqing@nimte.ac.cn
b) Corresponding authors. Qing Zhang, E-mail address: eqzhang@ntu.edu.sg

In all normal field-effect transistors, the channels are lightly doped and the device fabrication is always based on the formation of junctions between the channels and source/drain contacts.[1] Recentlly, junctionless Si nanowire transistors have been demonstrated, where the channel is of homogeneous doping polarity and uniform doping concentration across the channel without any junction formation.[2,3] However, these junctionless transistors are Si-based devices and strongly rely on the stringent requirements, such as silicon-on-insulator wafers, ultrathin gate dielectric, precise multistep photolithography, etc. In other words, fabrication of these junctionless FETs is still rather challenging. At the same time, dual-gate thin-film transistors (TFTs) have attracted intensive attention in chemical and biological sensor applications due to a flexible operation and easy adjustment of threshold voltage($V_{th}$).[4-6] In a typical dual-gate TFTs, the active channel is usually sandwiched between the up and bottom dielectric layer/gate electrodes.[7] Although these reported dual-gate TFTs demonstrate satisfactory device performance, complicated fabrication processes, say twice gate/dielectric deposition and precise photolithography steps are still required.[4] Therefore, a simple process is especially required to be developed to fabricate the junctionless dual-gate TFTs.

Compared to glass and plastic substrates, paper is an intriguing alternative to fulfill low-cost demands, because it is ubiquitous in daily life and by far the cheapest flexible substrate.[8] As paper surface is rough and porous, it is a challenging task to fabricate TFTs on paper substrates with high performance. Up to now, several groups have reported organic TFTs,[9,10] oxide-based TFTs[11,12] on paper substrates. However, the device performances and stabilities are still far from satisfaction. For example, the operation voltages of these paper TFTs are usually higher than 10 V, which is too high to be used for portable applications. Thus, novel gate dielectric materials that offers both low-temperature processability and high gate specific capacitance are highly required for paper TFT fabrication. Recent reports on polymer electrolytes [13,14] and inorganic nanogranular electrolytes[15] with huge electric-double-layer (EDL) capacitances have been attracted a lot of attention to this challenging issue.

In this paper, dual in-plane-gate electrostatic modulation of the self-assembled junctionless oxide channel by the solution-processed chitosan films are experimentally demonstrated on paper substrates. Our results also show that the threshold voltage ($V_{th}$) can be effectively tuned from -0.14 V ~ 1.15 V when the second in-plane gate bias switches from 2.0 V to -2.0 V. Such dual-gate organic/inorganic hybrid paper TFTs exhibit a large field-effect mobility (>10 cm$^2$/Vs), a high current on/off ratio (>10$^6$) and a small subthreshold swing (<100 mV/decade). At last, OR logic gate was experimentally demonstrated on such low-voltage dual-in-plane gate paper thin-film transistors.

Fig. 1(a) and (b) show the schematic pictures of the junctionless paper TFTs gated through solution-processed chitosan with one-in-plane and dual-in-plane gate architectures. The most attractive features of such TFT fabrication processes can be summarized as follows: (i) The channel and all in-plane electrodes (source, drain, in-plane gates) can be deposited and patterned by one-step magnetron sputtering process with only one nickel shadow mask at room temperature. (ii) Compared to the conventional dual-gate TFTs with top and bottom gate electrodes, such dual-in-plane gate junctionless TFTs can be easily fabricated. It is interesting to note that our TFTs are gated through two capacitantors (C1 and C2) coupled by the bottom ITO film (marked in Fig 1(a) in red). Thus, the gate capacitance should result from the series connection of the two capacitors. The low-voltage operation of the TFTs is due to the formation of electric-double-layer, as shown in Fig. 1(c). Dry chitosan film has a low ionic conductivity of $10^{-9}$ Scm$^{-1}$. Although there are abundant amino and hydroxyl functional groups, the groups are grafted on the polysaccharide framework with strong chemical bonds and can not move as mobile ions. However, when water is incorporated, some free amino groups in the chitosan backbone are partly protonated ($-NH_2+H_2O \leftrightarrow NH_3^+ +OH^-$) leading to the formation of hydroxide ions. A higher ionic conductivity of $10^{-5}$ Scm$^{-1}$ can be achieved due to the mobile OH$^-$ anions.[16] In this work, acetic acid is used to modulate the ionic conductivity through its mobile protons to neutralize OH$^-$ anions.[17] The excess CH$_3$COO$^-$ and protons move in response to electric field applied by the gate to form EDL capacitance. Similar to other electrolytes, protons are repelled to the interface between the chitosan layer and back ITO when a positive gate voltage is applied and form a Helmholtz layer, where equal density of charges are induced. The motion of protons could be a hopping process from one oxygen atom to another with the hydrogen bonds.[18] The major advantage of the EDL effect in chitosan dielectric is that the specific capacitance is extremely large at low frequencies so that a small gate voltage variation (~1.5 V) could induce several orders of magnitudes' change in the drain current. Fig. 1(d) shows a photo of the TFT arrays fabricated on a flexible paper substrate.

Figure 2 (a) shows the cross-sectional SEM image of the solution-processed chitosan film on silicon substrate. The thickness of the chitosan film is estimated to be about 6.0 μm. As chitosan is a proton conductor, it displays dipolar, ionic and electrolyte/electrode interfacial related relaxations in different frequency regions.[19] Fig. 2(b) shows the specific capacitance-frequency curve of a ITO/chitosan/ITO capacitor with 6.0 μm-thick chitosan dielectric. The specific capacitance decreases rapidly at high frequencies, levels out towards a plateau value at intermediate frequencies, and reaches to the maximum value of 3.3 μF/cm$^2$ at 20 Hz. According to the value of the ionic conductivity and phase angle, as shown in Fig. 2(c), there are two different frequency regions where the ionic species affect the frequency response differently. At lower frequencies ($f$ <150 Hz for $\theta(f)$ <–45$^\circ$), most of the ionic species accumulate near the electrode/chitosan interfaces, thus, an apparent drop in ionic conductivity.[13] This capacitive behaviour is dominated by an EDL formed at the ITO/chitosan interfaces. At higher frequencies ($f$ >150 Hz for $\theta(f)$ >–45$^\circ$), the ionic conductivity of the chitosan increases with frequencies. This capacitive behaviour is considered to be resulted from the dipolar relaxation of the chitosan dielectric. The impedance spectroscopy of the chitosan dielectric is similar to the typical solid electrolytes and ion gels, indicating that it is an electronically insulating, ionically conducting dielectric.[20,21] To further investigate the dielectric properties of the chitosan electrolyte, Fig. 2 (d) shows the gate leakage current curve with a small hysteresis window, which could be due to the different transport mechanisms of the positive ions and negative ions in the chitosan film under the influence of external electric field.[22] The leakage current is less than 80 pA at 1.5 V, seven orders of magnitude smaller than the channel current ($I_{ds}$), indicating that $I_{ds}$ is not be affected by the leakage current. The small leakage current suggests that electrochemical reactions and ion currents are negligible in such electrolyte dielectrics.[23, 24]

Figure 3 (a) shows the typical output curves of the junctionless paper TFTs with normal bottom ITO gate. It reveals good current saturation behaviors at high $V_{ds}$ and excellent linear characteristics of $I_{ds}$ at low $V_{ds}$, suggesting that low resistance of ohmic contacts are formed between the ITO channel layer and ITO source/drain electrodes. Fig. 3(b) shows the corresponding transfer characteristics in the dark and under visible light. In the dark, the device shows an on/off current ratio ($I_{on/off}$) of $4 \times 10^6$ and a subthreshold slope (S) of 80 mV/decade. The off current is increased slightly under the visible light irradiation, and it is reversible when the light is tuned off. The electron field-effect mobility ($\mu$) of the device in the saturation operation regime can be calculated using $I_{ds}=[(WC_i)/(2L)]\mu(V_{gs}-V_{th})^2$, where L is the channel length, W is the channel width, $C_i$ is the specific gate capacitance at 20 Hz, $V_{th}$=0.58 V, the threshold voltage is calculated from x-axis intercept of $(I_{ds})^{1/2}$-$V_{gs}$ plot. Using this model, the field-effect electron mobility was estimated to be ~10 cm$^2$/Vs. A very small hysteresis window of 0.05 V was observed in the $I_{ds}$-$V_{gs}$ curves, which is likely due to the mobile protons in the chitosan dielectric. The proton density is calculated to be $1.8 \times 10^{12}$ /cm$^2$ based on the equation of N=$\triangle V_{th}C_i$/e, where $\triangle V_{th}$=0.09 V is the threshold voltage shift between the dual sweeping. Bias stress stability of such unpassivated paper TFTs was also investigated at $V_{ds}$=1.5V and $V_{gs}$=-0.5 V for 1000 s in the dark, as shown in Fig. 3(c). No detactable threshold voltage shift was observed, suggesting good stability of the device at air ambient. In order to find out whether electrochemical doping existed in the ITO channel during the device operation, low-frequency pulse respond characteristic was employed using a square-shaped $V_{gs}$ with pulsed amplitude of -0.5V to 1.5 V, as shown in Fig. 3(d). The device maintains a current on/off ratios of ~$10^7$ and no decreasing in current is observed. Since $I_{ds}$ would not return to its original value after gate scanning when chemical doping or a chemical reaction had occurred,.[17] Our results here indicate that no chemical doping or chemical reaction occurs at the chitosan/ITO interface under the given gate potential.

Fig. 4 (a) shows the transfer curves of the dual-gate junctionless TFTs on paper substrates in the saturation regime ($V_{ds}$=1.5 V) with an in-plane-gate voltage from 2.0 V to -2.0 V. All transfer curves were sweeping from the negative to positive gate voltage. When $V_{G2}$ is swept from 2.0 V to -2.0 V, the transfer curves systematically shift from the left to the right. The threshold voltage ($V_{th}$) modulation by the second in-plane-gate voltage bias of $V_{G2}$ can be explained as follows. A negative $V_{G2}$ partially depletes electrons in the channel. To compensate the depletion, the more positive gate voltage bias is needed, which makes the $V_{th}$ move toward positive direction. In contrast, a positive $V_{G2}$ accuminates electrons in the channel, which creates an additional current and effectively shifts the $V_{th}$ toward negative direction. Fig. 4 (b) shows the $(I_{ds})^{1/2}$ versus $V_{bottom-gate}$ transfer curves with different $V_{in-plane-gate}$. The threshold voltage $V_{th}$, determined by the intercepts lines with the $V_{bottom-gate}$ axis, is adjustable from 0.07 V to 0.88 V. These results indicate that an effectively electrostatic coupling is realized between the in-plane-gate and the ITO channel. For dual-gate structures, the field-effect mobility is improved to ~17.6 cm$^2$/Vs. By the way, the subthreshold swing of ~80 mV/decade and a current on/off ratio of ~4×10$^6$ are highly repeatable at different in-plane-gate voltage biases.

Fig. 5(a) shows the transfer curves (in saturation region) of the dual-in-plane gate junctionless paper TFTs at $V_{ds}$=1.5 V for different voltage biases of the secondary gate ($V_{G2}$=2.0, 1.0, 0, -1.0, -2.0 V). Positive $V_{G2}$ makes the transfer curve shift to the negative direction, whereas negative $V_{G2}$ makes the transfer curve shift to the positive direction. Saturation current is systematically increased by increasing the $V_{G2}$. The $V_{th}$ of dual in-plane gate TFTs with different $V_{G2}$ was calculated from x-axis intercept of the square root of $I_{ds}$-$V_{gs}$ plot in Fig. 5(b). Such a dual in-plane-gate TFTs has a large adjustable $V_{th}$ range from -0.14 V to 1.15 V by tuning $V_{G2}$ from 2.0 V to -2.0 V, corresponding to depletion-mode and enhancement-mode, respectively.

Fig. 6 (a) shows the optical image of the fabricated device. To meet the requirement for the OR operation, the device structure was designed in a symmetrical manner. As seen in the figure, both gates are almost identical. To demonstrate the OR operation in a time domain, voltage pulse sequences with different duration times and periods were applied to the two gates, and the output current is shown in Fig. 6 (b). In the figure, 1.5 V and 0 V are defined as the high and low levels for the input, respectively. When both gates are in the low level (0 V), the drain current is blocked (off). When one gate is in the high level (1.5V), the drain current flows (on). This is exactly OR operation of the dual-gate TFTs. The device maintains a current on/off ratios of ~$10^6$ and no current decreases are observed.

The concept of the OR function implementation is shown in Fig. 7. In the figure, 1.5 V and 0 V are defined as the high ("1") and low ("0") levels for the input, respectively. The OR functionality can be implemented by controlling the overlap of the two depletion regions formed under different biases.[25] As shown in Fig. 7, when both gates are at "0", the depletion regions, created by two gates, fill at least the narrowest region of the channel completely, so the current is blocked (off). However, when any of the gates is "1", the TFT switches to the ON state, which results in an OR functionality.

In conclusion, we have demonstrated that self-aligned junctionless semiconducting oxide channels could be low-voltage electrostatic modulated by solution-processed chitosan-based proton conductors on paper substrates in a dua -in-plane gate figure. The operation mode of such junctionless paper transistors could be effectively tuned from depletion mode to enhancement mode by a second in-plane gate. An OR logic gate function was experimentally demonstrated. Such self-assembled junctionless paper transistors are promising for portable paper electronics.

**Acknowledgments**

**The authors are grateful for the financial supports from the National Program on Key Basic Research Project (2012CB933004), the National Natural Science**

**Foundation of China (11174300), and the Fok Ying Tung Education Foundation (Grant No. 121063).**

**Experimental Section**

Paper substrates used for oxide-based TFT arrays fabrication were commercially available papers for ink-jet printing. Radio-frequency (RF) magnetron sputtering and plasma-enhanced chemical vapor deposition (PECVD) were used for the fabrication of paper TFTs gated by solution-processed chitosan dielectrics. The entire process was performed at room temperature. First, a 2.0-μm-thick nanogranular $SiO_2$ film was deposited on paper substrates for surface passivation and smoothness improvement. Second, a 200-nm-thick indium-tin-oxide (ITO) film was deposited on passivated paper substrates by RF magnetron sputtering at 0.5 Pa. Third, chitosan solution (2 wt% in acetic acid) was spined onto the ITO/$SiO_2$/paper substrates and dried in air ambinet. At last, the fabrication of the oxide-based TFT arrays on paper substrats was completed by RF sputtering of ITO source/drain and dual in-plane gate electrodes through a nickel shadow mask with a channel width to length of 1000/80 μm. Here we should point out that a thin ITO active channel can be self-aligned between two source/drain electrodes during sputtering deposition. The structural characterization of the chitosan-based film was performed by scanning electron microscope (Hitachi S-4800 SEM). Electrical characterizations of chitosan dielectric and the paper TFTs were performed by an impedance analyzer (Agilent 4294A) and a semiconductor parameter analyzer (Keithley 4200 SCS) at room temperature in air ambient.

**Figure captions**

Figure 1. (a) and (b) Schematic pictures of the dual-gate paper TFTs gated by solution-processed chitosan dielectric. (c) Schematic picture of the low-voltage operation mechanism due to EDL formation. (d) Picture of the blended paper TFT arrays, indicating its flexible nature.

Figure 2. (a) Cross-sectional SEM image of the solution-processed chitosan film. (b) Capacitance-frequency curve of the chitosan dielectric in the frequency range from 20 to $10^6$ Hz. (c) Frequency dependent ionic conductivity and phase angle of the chitosan dielectric. (d) Gate leakage current curve of the chitosan dielectric.

Figure 3. (a) Output and (b) transfer characteristics of the paper TFTs gated by solution-processed chitosan dielectric. (c) The evolution of transfer curves of devices as a function of the applied -0.5 V negative bias stress time in the dark. (d) Low-frequency pulse respond characteristic of the device with $V_{gs}$ = -0.5 to 1.5 V and $V_{ds}$ = 1.5 V.

Figure 4. (a) The transfer characteristics of the dual-gate paper TFTs in the saturation regime ($V_{ds}$=1.5 V) at different in-plane-gate voltage biases. (b) The evolution of the $(I_{ds})^{1/2}$ versus $V_{bottom-gate}$ transfer curves of the same device.

Figure 5. (a) The transfer characteristics of the dual in-plane-gate paper TFTs in the saturation regime ($V_{ds}$=1.5 V) with different $V_{G2}$ ranging from 2.0 V to -2.0 V. (b) The $(I_{ds})^{1/2}$ versus $V_{gs}$ curves.

Figure 6. (a) Optical image of the fabricated OR device. (b) Current output in response to the two inputs applied as a pulse sequence. The OR operation with a large on/off ratio is obtained.

Figure 7. Schematic diagram of OR functionality operation. The depletion regions (in red) formed under different bias conduction.

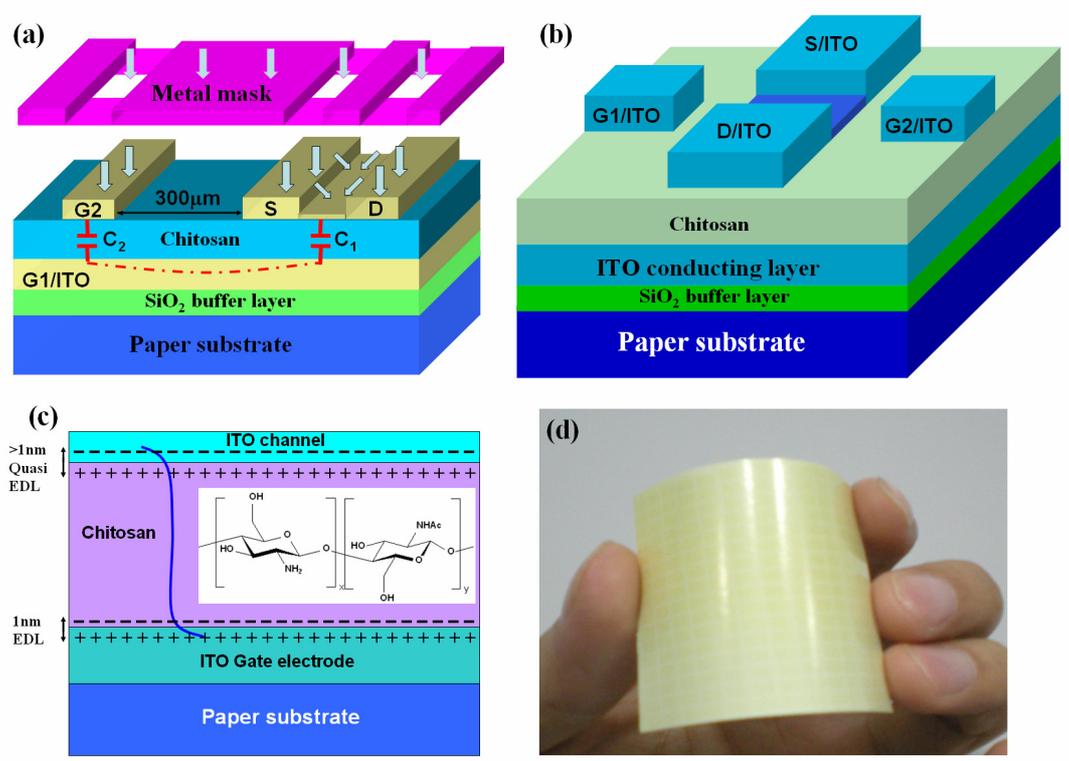

Fig.1

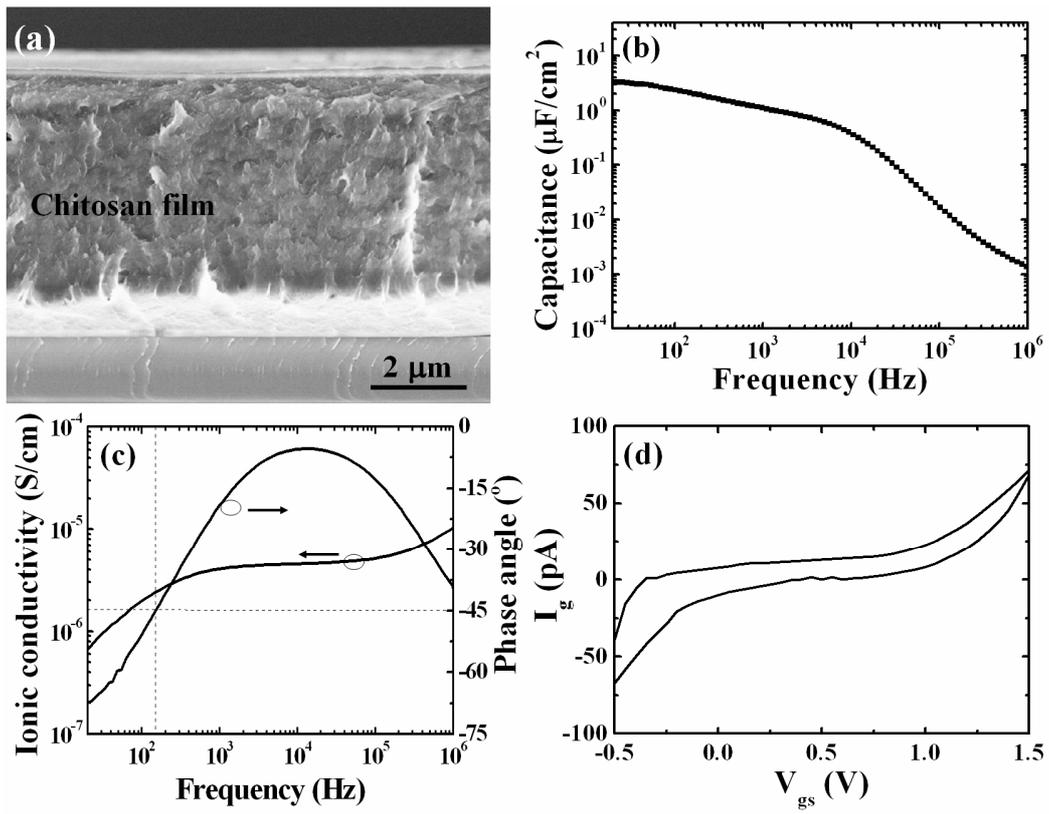

Fig.2

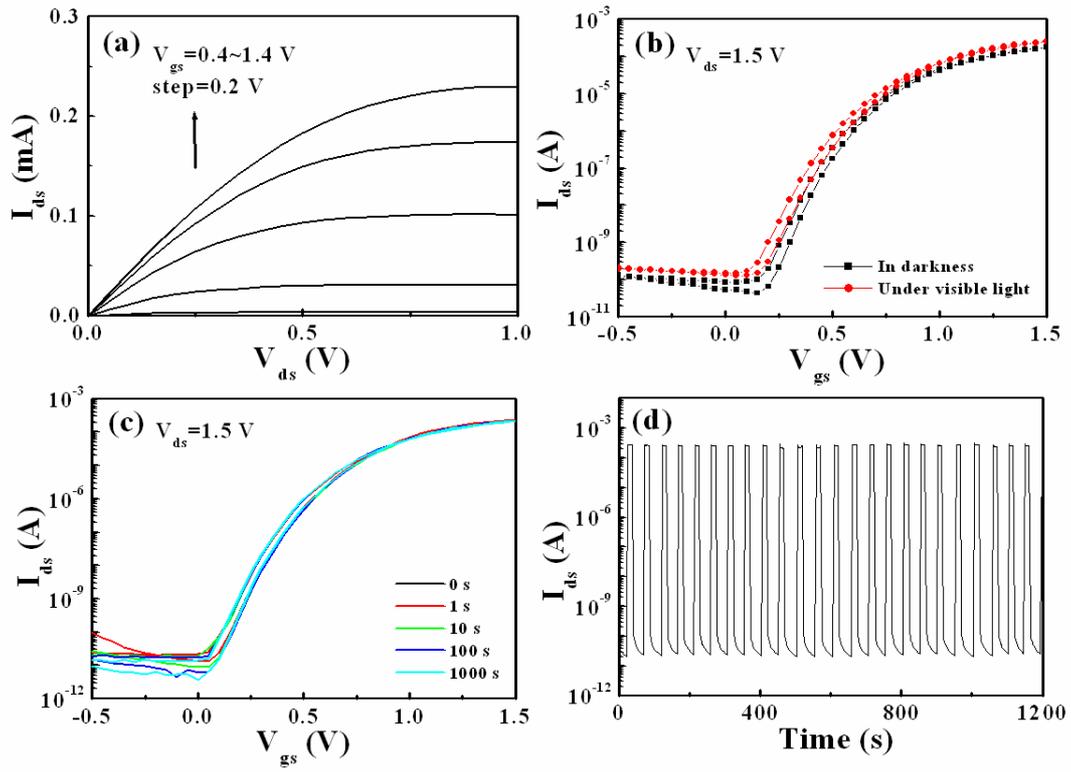

Fig.3

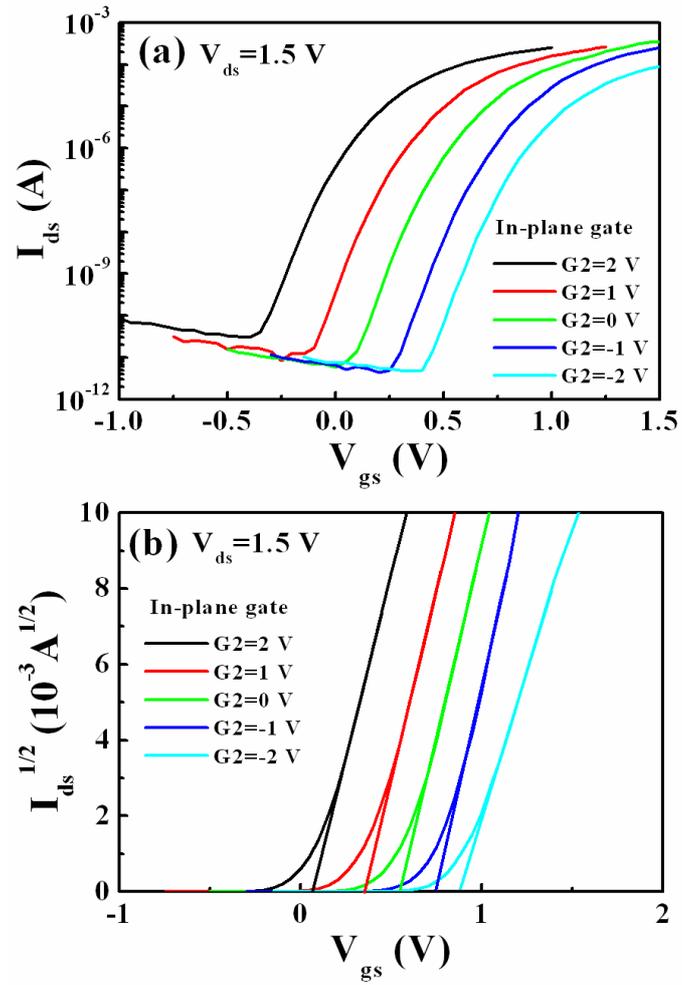

Fig.4

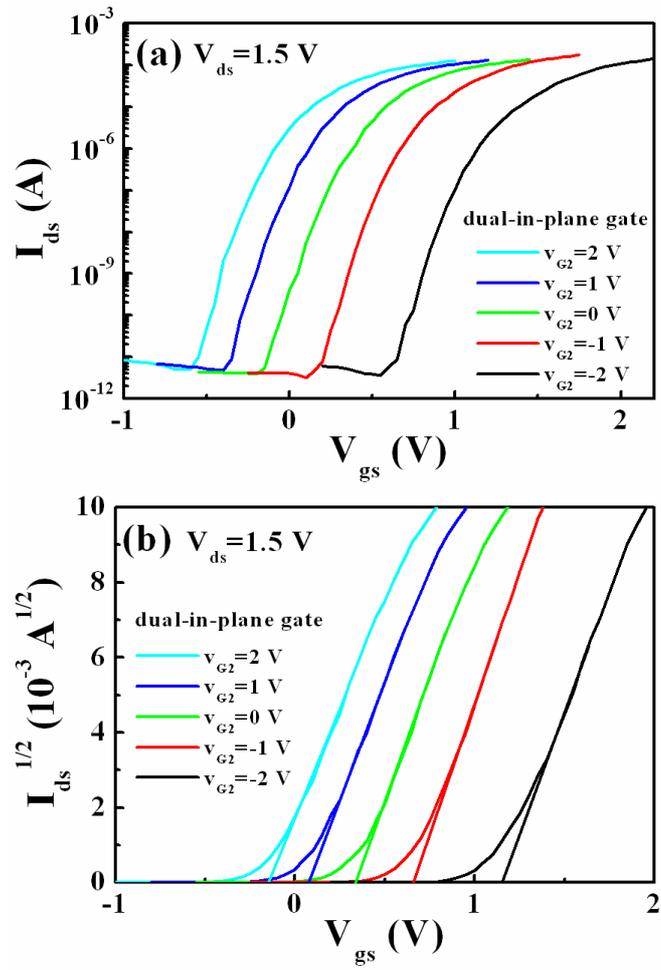

Fig.5

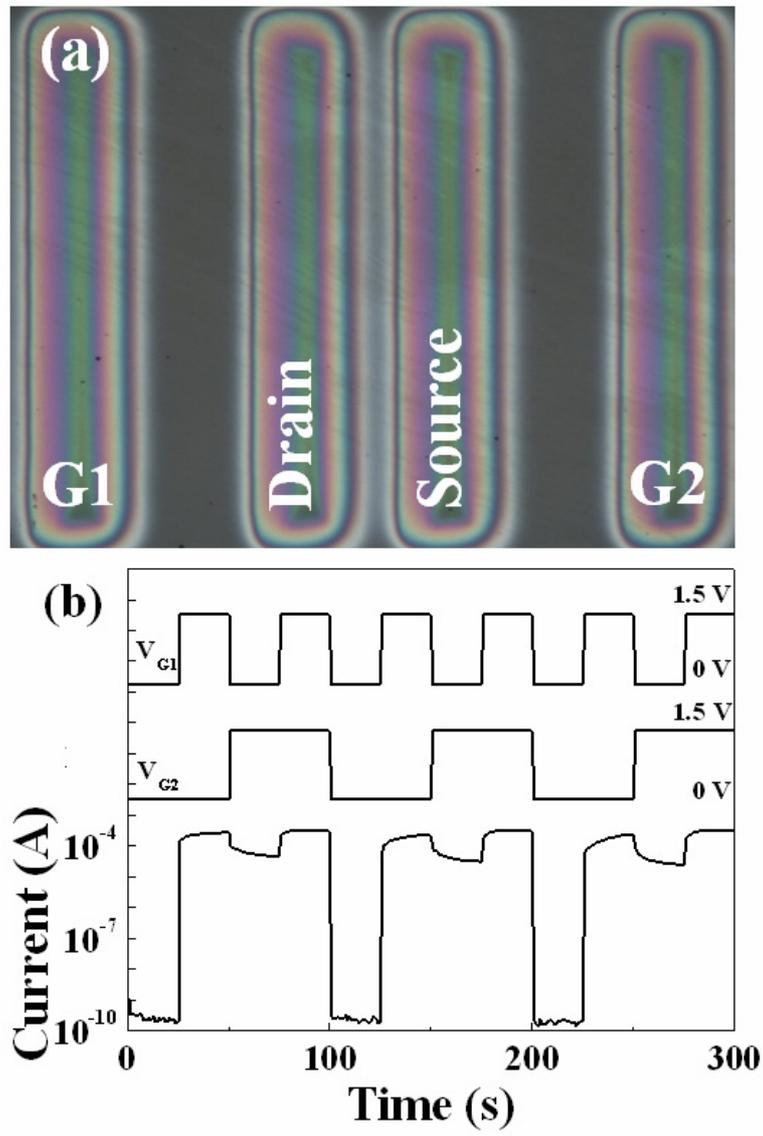

Fig.6

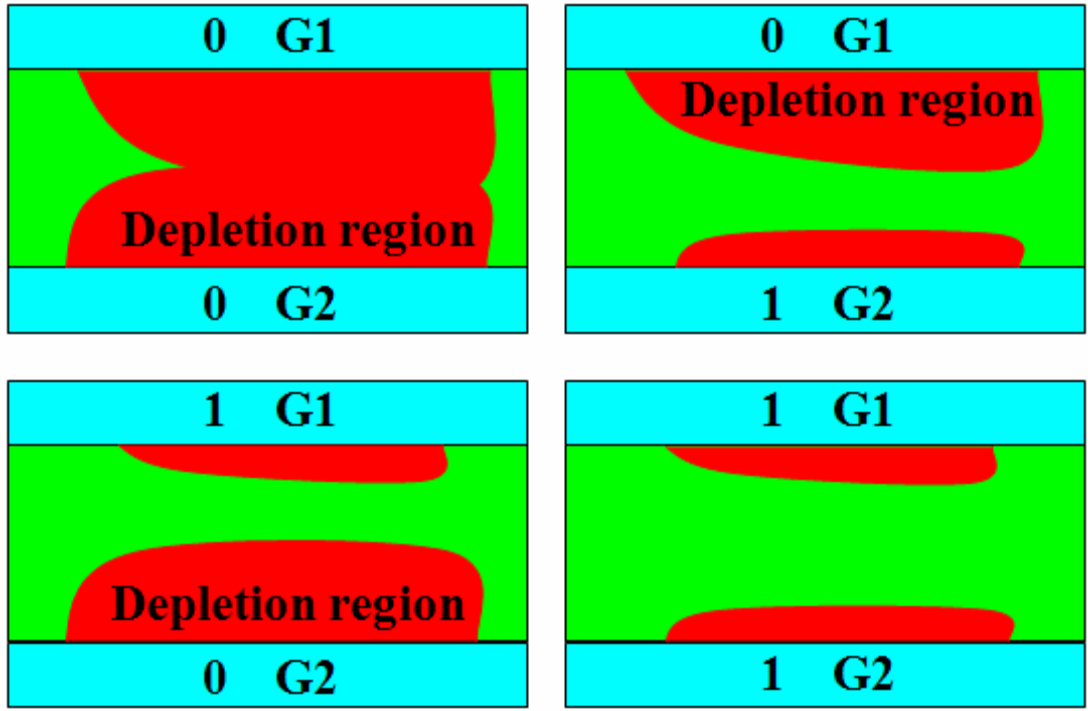

Fig.7